\documentclass[twocolumn,prl,a4paper,showpacs,floatfix,preprintnumbers,amsmath, amssymb,superscriptaddress]{revtex4}

\usepackage{amsmath}
\usepackage{graphicx}
\usepackage{subfigure}
\usepackage{psfrag}
\usepackage[centerlast]{caption}
\usepackage[dvips]{color}

\usepackage{graphicx}
\usepackage{dcolumn}
\usepackage{bm}

\renewcommand{\frac}[2]{\genfrac{}{}{.6pt}{}{#1}{#2}} 
\newcommand{\fracp}[2]{\genfrac{}{}{.3pt}{}{#1}{#2}} 

\newcommand{\be}{\begin{equation}}
\newcommand{\ee}{\end{equation}}

\newcommand{\nt}{N_{T}}
\newcommand{\ns}{N_{S}}
\newcommand{\nj}{N_{j}}
\newcommand{\tf}{t_{f}}
\newcommand{\U}{\mathcal{U}}

\newcommand{\ej}{\epsilon_{j}}
\newcommand{\cj}{\hat{c}_{j}^{\phantom{\dag}}}
\newcommand{\cjd}{\hat{c}_{j}^{\dag}}

\newcommand{\pj}{\phi_{j}}


\begin{document}


\title{Macroscopic Superpositions of Phase States with Bose-Einstein Condensates}

\author{F. Piazza}\affiliation{CNR-INFM BEC center and Dipartimento di Fisica, Universit\`a di Trento, I-38050 Povo, Italy}
\author{L. Pezz\'e}\affiliation{Institut d'Optique, Campus Polytechnique, F-91127 Palaiseau, France}
\author{A. Smerzi}\affiliation{CNR-INFM BEC center and Dipartimento di Fisica, Universit\`a di Trento, I-38050 Povo, Italy}

\date{\today}
                      
\begin{abstract}
Quantum superpositions of macroscopically distinguishable states having distinct phases
can be created with a Bose-Einstein condensate trapped in a periodic potential.
The experimental signature
is contained in the phase distribution of the interference patterns
obtained after releasing the traps. 
Moreover, in the double well case, this distribution exhibits a dramatic dependence
on the parity of the total number of atoms. 
We finally show that, for single well occupations up to a few hundred atoms, 
the macroscopic quantum superposition can be robust enough against decoherence to be experimentally revealable 
within current technology.

\end{abstract}

\maketitle

{\it Introduction}.
Nonlinearity is crucial for the creation of a superposition of 
macroscopically distinguishable states, often referred to in the literature as ``Schr\"odinger's cat'' 
\cite{Yurke_1986,Milburn_1986,Sanders_1992,Ourjoumtsev_2007}.
It was first suggested by Yurke and Stoler \cite{Yurke_1986}  
to use a kerr medium to create a superposition of photonic coherent states having different phases
and to detect it via homodyne interferometry.
However, it was quickly realized that the experimental realization of these special states is rather prohibitive due to the typically 
weak optical nonlinearities and strong losses in the medium. 
Recent advances in atomic physics, where 
large nonlinearities arise naturally, are changing this scenario and encouraging a renewal of early proposals. 
Macroscopic superpositions with a small number
of trapped ions have been created in \cite{Leibfried_2005}, 
while Bose-Einstein condensates (BECs) might provide
even brighter sources thanks to their 
inter-particle interaction which introduce a large 
kerr-like coupling of matter waves.
In particular, several thermodynamical and dynamical schemes for creating
a superposition of two BEC states differing by a macroscopically large
number of particles have been proposed in the literature
\cite{noon_history}. These so called
``NOON states'' maximize particle entanglement and can be useful in
quantum information protocols as, for instance, Heisenberg-limited
interferometric phase estimation.

In this manuscript we discuss an experimentally feasible protocol for the creation and detection of a macroscopic superposition
of states having different relative phases with a Bose-Einstein condensate trapped in a periodic potential.
These entangled states \cite{Yurke_1986}
are generated by the nonlinear unitary evolution governed
by the decoupled $\ns$-mode Bose-Hubbard Hamiltonian: 
$\hat{H}_{\ns}=\sum_{j=1}^{\ns} \ej\cjd\cj+\fracp{\U_j}{2}\cjd\cjd\cj\cj$ and are closely related, 
as will be shown later, to NOON states.
Here $\cj$ ($\hat{c}_j^{\dag}$) annihilate (create) 
a particle in the $j^{\mathrm{th}}$ condensate, 
$\ej$ is the energy offset due, for instance, to an external potential superimposed to the optical lattice and
$\U_j$ is the single-condensate interaction energy. 
For a double well system, $\ns=2$, we show, by simulating the formation of several single-shot 
interference density profiles with a many-body
Montecarlo technique \cite{Javanainen_1996}, that the relative phase probability
distribution contains distinct peaks. Each peak corresponds to a phase state component of 
the superposition and its position depends dramatically on the parity of the total number of atoms. 
We demonstrate that a clear signature of the creation of these macroscopic superpositions also appears 
in the interference of an array of BECs, even in a single-shot density profile.
We finally account for the problem of decoherence due to one, 
two, and three-body losses \cite{Sinatra_1998}. 
For typical experimental trapping parameters and single well occupations up to 
a few hundred atoms, the decoherence time can be about $500$ ms, 
longer than the typical formation time of the macroscopic superposition of phase states which can therefore 
be experimentally created and detected within current technology.

\begin{figure}[]
\begin{center}
\subfigure[]{
             \includegraphics[scale=0.43]{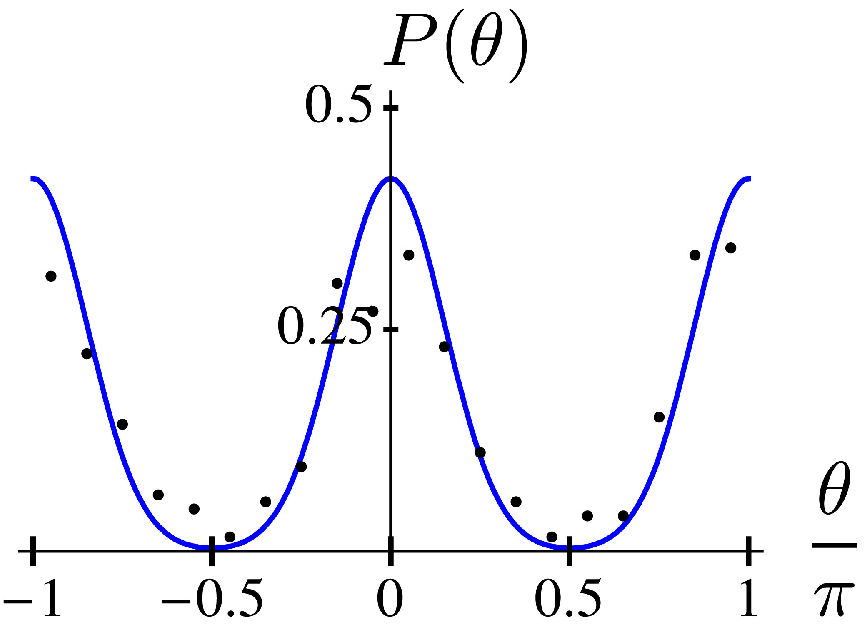}}
\hspace{0.0in}
\subfigure[]{
             \includegraphics[scale=0.43]{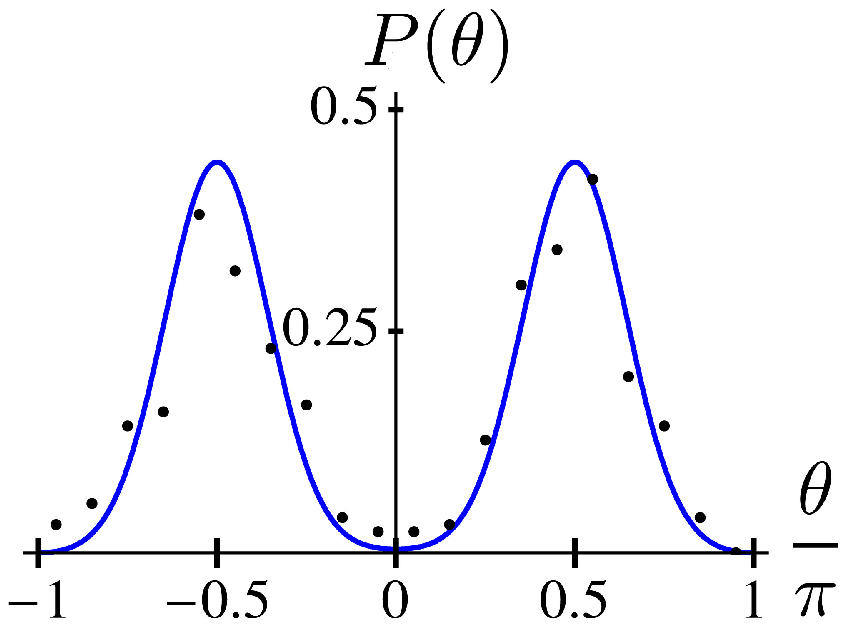}}\\
\vspace{0in}
\subfigure[]{\includegraphics[scale=0.21]{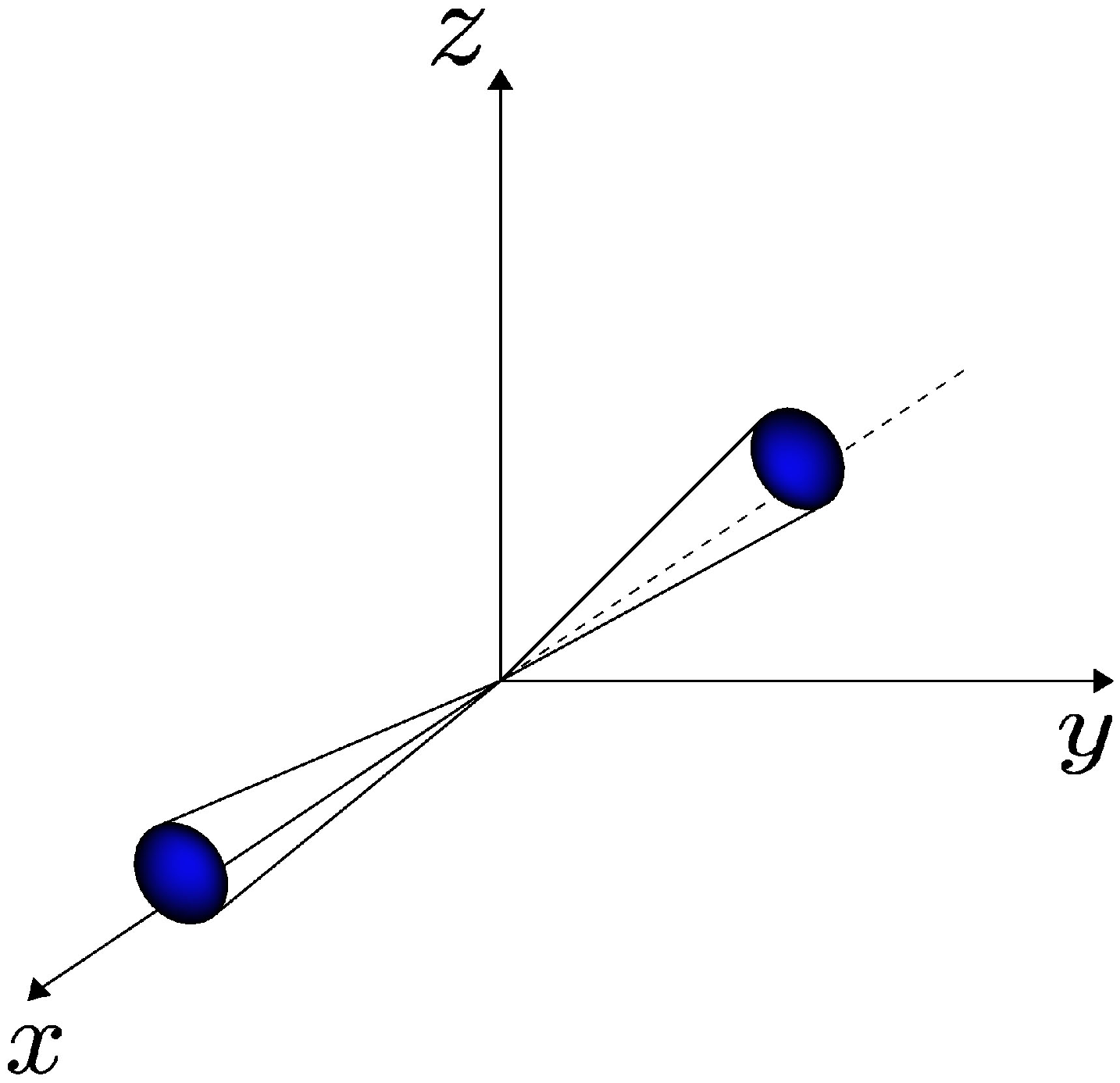}}
\hspace{0.25in}
\subfigure[]{\includegraphics[scale=0.21]{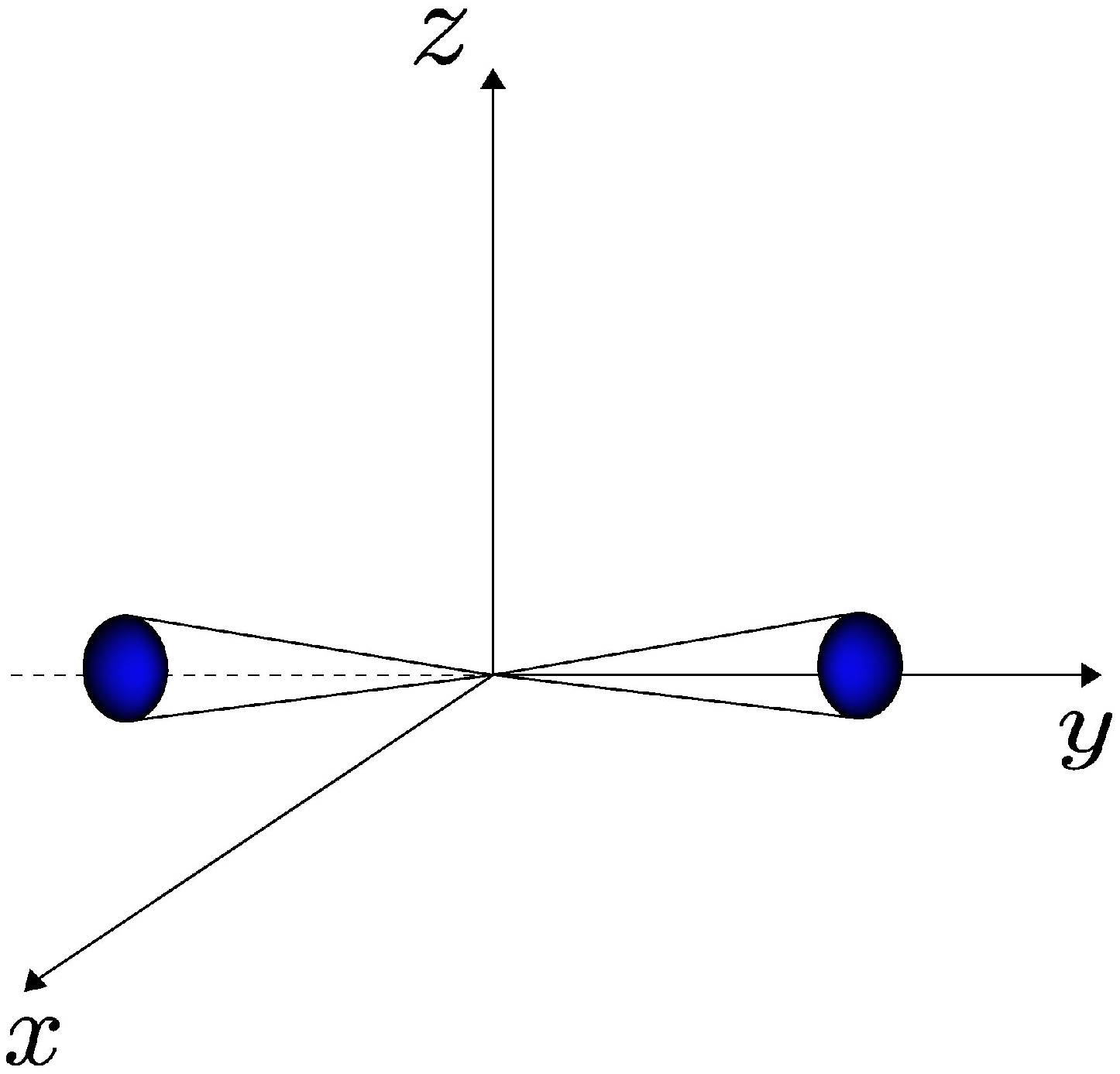}}%
\caption{\small{(color online). a),b) Montecarlo phase distribution (dots, with the phase values distributed over 20 bins)
and the SU(2) phase distribution $P(\theta)$ (solid line), for the
state Eq.(\ref{superp}) with $s=1.5$. In (a) $N_T$ is even and
$P(\theta)\sim\cos^{2 N_T}(\theta)$; in (b) $N_T$ is odd and
$P(\theta)\sim\sin^{2 N_T}(\theta)$.
c),d) Schematic pseudo angular momentum representation of Eq.(\ref{superp}) for even (c) and 
odd (d) values of $\nt$.} \label{simul}}
\end{center}
\end{figure}

{\it Double well potential.} We first consider a BEC trapped in a symmetric double 
well potential ($\epsilon_1=\epsilon_2$, $\U_1=\U_2\equiv \U$)
and rewrite the Hamiltonian in the pseudo angular momentum 
representation, $\hat{H}_{2}=\U \hat{J}_z^2$ \cite{nota02,noon_history}.
As initial condition we choose a state of $N_T$ particles
$|\psi(\theta_0)\rangle=\sum_{n=0}^{N_{T}} C_{n}e^{-in\theta_0}|N_{T}-n,n\rangle_z$ 
(where $|N_{T}-n,n\rangle_z$ is a number Fock state \cite{nota02b}).
The state $|\Psi(t_{\fracp{\pi}{2}})\rangle=e^{-i \hat{H}_2 t_{\pi/2}/\hbar} |\psi(\theta_0)\rangle$, 
obtained after a time $t_{\pi/2}\equiv\frac{\hbar}{\U}\frac{\pi}{2}$, can be formally written as 
\begin{equation} \label{superp}
|\Psi(t_{\fracp{\pi}{2}})\rangle=
\fracp{e^{-i\fracp{\pi}{4}}}{\sqrt{2}} |\psi(\theta_0+\fracp{\pi}{2}\xi)\rangle +
\fracp{e^{+i\fracp{\pi}{4}}}{\sqrt{2}} |\psi(\theta_0-\pi+\fracp{\pi}{2}\xi)\rangle,
\end{equation}
where $\xi=0$ ($\xi=1$) for $\nt$ even (odd) \cite{nota03}. 
It is instructive to project the state Eq.\eqref{superp} over the SU(2) basis states 
$|N_T,\theta\rangle=\frac{1}{\sqrt{2\pi}\sqrt{N_T+1}} \sum_{n=0}^{N_{T}}e^{-in\theta}|N_{T}-n,n\rangle_z$ \cite{Vourdas_1990}.
If the initial $|\psi(\theta_0)\rangle$ has a relative phase distribution localized about $\theta_0$, 
the phase distribution of the evolved state,
$P(\theta) = |\langle N_{T},\theta |\Psi(t_{\pi/2}) \rangle|^{2}$,  
is characterized by $2$ peaks, see Fig.~\ref{simul}, 
which is the consequence of being $|\Psi(t_{\pi/2}) \rangle$ a
superposition of $2$ states having different relative phases.
Since the peaks have a finite width (see discussion below),  
the highest visibility is reached just at $t_{\pi/2}$, 
when the peaks are maximally separated. 
For instance, when the initial state is given by a binomial distribution,
$C_n=\frac{1}{2^{N_T/2}}\sqrt{\binom{N_T}{n}}$ and $\theta_0=0$, 
the two components of the superposition are exactly orthogonal
$\langle \psi(\fracp{\pi}{2}\xi)| \psi(-\pi+\fracp{\pi}{2}\xi)\rangle = 0$, 
and $P(\theta)$ has peaks of width $w\sim 1/\sqrt{N_T}$.
With trapped BEC, a realistic scheme for the creation of a
superposition of two states having different relative phases involves the sudden splitting
of a single condensate \cite{nota04}.
The BEC is left in a state slightly squeezed in the relative number of particles, 
$|\psi_0\rangle \equiv |\psi(\theta_0=0)\rangle\sim \sum_{n=0}^{N_{T}}e^{-(n-N_T/2)^{2}/4\sigma_{s}^{2}}|N_{T}-n,n\rangle_z$, 
which provides the initial condition of the decoupled non-linear evolution.
The width of the relative number distribution is $\sigma_{s}=\sqrt{N_{T}}/(2s)$ and $s$ is the squeezing parameter. 
After a time $t_{\pi/2}$, this state evolves in the superposition Eq.(\ref{superp}) 
(for moderate initial number-squeezing the overlap between the two components is exponentially small).
We release the confining potential and let the condensate ballistically expand and overlap,
giving rise to an interference pattern  \cite{Andrews_1997,Oberthaler_2006}
from which we extract a single value of the relative phase.
We will show that the phase distribution obtained upon several
interference experiments
is reasonably well described by the SU(2) probability $P(\theta)$, when the initial state is number-squeezed $s > 1$.
Let us consider a simple model which can be numerically studied 
with a Montecarlo technique \cite{Javanainen_1996}.
First, we simulate the formation of a single-shot interference density profile.
The probability to detect $N_T$ particles at the same time 
and positions  $\{x_{1},x_{2},...,x_{N_T}\}$ is 
\begin{equation} \label{manyb}
P_{\{x\}}=
\frac{\langle\Psi(t_{\fracp{\pi}{2}})|\hat{\phi}^{\dag}(x_{1})...\hat{\phi}^{\dag}(x_{N_T})\hat{\phi}^{}
(x_{N_T})...\hat{\phi}^{}(x_{1})|\Psi(t_{\fracp{\pi}{2}})\rangle}{N_T!}.
\end{equation}
Here $\hat{\phi}(x)=\phi_{1}(x)\hat{c}_1+\phi_{2}(x)\hat{c}_2$, where $\phi_{j}(x)$ is a 
normalized wave function of the $j^{th}$ condensate. 
The numerical calculation is simplified by using two counter-propagating 
plane waves: $\phi_{1}(x)=\phi_{2}(x)^{*}=e^{i\pi x}$, with $x\in [0,1]$ . 
After generating $N_T$ random positions distributed with $P_{\{x\}}$,
the phase $\theta$ is extracted by fitting the density profile to the function 
$\rho(x;\theta)=1+\cos(2\pi x+\theta)$.
We repeat several times the interference protocol to obtain a probability phase distribution.
In FIG.~\ref{simul}(a),(b) we show the results of 400 independent phase estimations
(dots) for $N_T=10$ (a) and $N_T=11$ (b).
The agreement with the SU(2) phase distribution (solid line) holds already for small values of $N_T$ and improves for a
larger number of atoms as long as $s > 1$.
Thus, the single-shot density profile can be approximated by
interfering spatial wave functions with 
a relative phase randomly sampled from $P(\theta)$.
We emphasize that this agreement is not obvious. In particular, SU(2) fails for 
phase squeezed states, $s \lesssim 1$, whose phase distributions are 
better matched by projecting over binomial phase states.   

In FIG. \ref{simul}(a),(b) we can distinguish  
two peaks separated by $\pi$, 
each corresponding to a different phase component of Eq.(\ref{superp}).
The phase shift due to a change in 
the parity of $\nt$ can be clearly seen as a shift of $\pi/2$ 
among the distributions of (a) and (b). As expected, the width of the 
peaks $w\sim s/\sqrt{N_T}$ increases with the relative number squeezing 
of the initial state. 
In typical experimental conditions there is no control on the parity of the total number of atoms. 
In this case, the system is described by a classical mixture of superposition states, 
half corresponding to odd $\nt$ and half to even $\nt$. 
For instance, the phase distribution at $t_{\pi/2}$ would show 4 peaks, with widths scaling as $\sim 1/\sqrt{N_T}$, thus clearly 
distinguishable already for small $N_T$.
Notice that the occurrence of two peaks (or four if the parity of $N_T$ is not controlled) 
in the phase distribution constitutes, by itself, a signature of the presence of a quantum superposition, rather than a statistical
(non coherent) mixture with the same components. Indeed, if the system had decayed into a mixture at any time $t < t_{\pi/2}$, 
it could not have evolved into a macroscopic superposition. This because the state Eq.\eqref{superp}
is created and lives in a very narrow temporal window, $\sim t_{\pi/2}\pm \hbar/\U N_T$ \cite{nota100}.

To conclude, we point out that, when the initial state is given by a binomial distribution, we can rewrite Eq.(\ref{superp}) as 
$|\Psi(t_{\fracp{\pi}{2}})\rangle=(e^{-i\fracp{\pi}{4}}|N_T,0\rangle_x+e^{i\fracp{\pi}{4}}|0, N_T\rangle_x)/\sqrt{2}$
( $|\Psi(t_{\fracp{\pi}{2}})\rangle=(e^{-i\fracp{\pi}{4}}|N_T,0\rangle_y+e^{i\fracp{\pi}{4}}|0, N_T\rangle_y)/\sqrt{2}$ )
for $N_T$ even (odd) \cite{nota02b}.
The representation of these states in the Bloch-sphere is shown in FIG.~\ref{simul}(c),(d).  
The even (odd) $|\Psi(t_{\fracp{\pi}{2}})\rangle$ are 
simply linked by a $\pi/2$ collective rotation about the $z$ axis,
and may 
be both transformed into a NOON state $(e^{-i\fracp{\pi}{4}}|N_T,0\rangle_z + e^{i\fracp{\pi}{4}}|0,N_T\rangle_z)/\sqrt{2}$ 
with a $\pi/2$ rotation 
about the $y$ ($x$) axis \cite{nota11, notamodi}.
Therefore, a macroscopic superposition of two binomial states carries the same amount of particle entanglement (which is conserved by 
local unitary transformations) of the NOON state \cite{Pezze_2007}.
As a consequence, it would be, for instance, as useful as the NOON
state for quantum information protocols. We notice that, in the same way,
a NOON state can be created with high fidelity from a
slightly number squeezed initial state.

{\it Lattice potential.} The protocol 
for creating phase cats in the double-well
has a direct generalization to an array of condensates.
We consider a superfluid BEC trapped in a one-dimensional optical lattice modulated by a harmonic potential. 
The initial state, a product of coherent or slightly squeezed states 
$\prod_{j=1}^{N_S} |s_{j}(0)\rangle$ \cite{Orzel_2001} localized in each lattice well,
is created by increasing the interwell barriers rapidly enough to prevent the system from reaching the
Mott insulating phase \cite{nota04}.
The state then evolves with the decoupled Hamiltonian $\hat{H}_{\ns}$:
\begin{equation}
\label{state_lattice}
|s_{j}(t)\rangle=C_{j}\sum_{n=0}^{+\infty}e^{-\fracp{(n-\nj)^{2}}{4\sigma_{s}^{2}}}
e^{-i\fracp{n\ej t}{\hbar}}e^{-i\fracp{n(n-1)\U_{j}t}{2\hbar}}|n\rangle,
\end{equation}
where $\nj$ is the mean occupation number of the $j^{\mathrm{th}}$ site, $C_{j}$ is a normalization constant, 
$\sigma_{s}=\sqrt{\nj}/s \ll N_T$, and $s$ is the squeezing parameter which depends on the lattice ramping time \cite{nota05}.
At time $t_{\pi}$ the lattice is switched off and an image of the cloud is taken 
after a time of flight $\tf$.
To construct the single-shot interference density profile 
$\rho(x;\theta_{0},\dots,\theta_{\ns-1})$, we sample the values of 
the phase of the wave-function in each well with the distribution 
$P_{j}(\theta) \propto |\langle\theta|s_{j}(t_{\pi})\rangle|^{2}$,
being $|\theta\rangle=\fracp{1}{\sqrt{2\pi}} \sum_{n=0}^{+\infty} e^{i n \theta } |n\rangle$.
A simple calculation yelds
$\rho(x;\theta_{1},\dots,\theta_{\ns}) \propto |\sum_{j=1}^{\ns}\sqrt{\nj}e^{i\theta_{j}}e^{\fracp{im(x-jd)^{2}}{2\hbar\tf}-\fracp{(x-jd)^{2}}{l^{2}}}|^{2}$, 
where $d$ is the lattice period, $m$ is the atomic mass,
$l=\hbar\tf/ml_{0}$ and $l_{0}$ is the initial width of the Gaussian wave function describing each well along the lattice direction.
Finally, we extract a single value for the phase by fitting the density profile to the function
$[1+\beta\cos(\theta+2\pi x/L)]G(x)$, where $\beta$ and $\theta$ are fitting parameters, 
$L=h\tf/md$ is the period of the first-harmonic modulation of the profile, and $G(x)$ is a 
Gaussian envelope that accounts for the finite dimensions of the system \cite{modulation}.

\begin{figure}[tbp]
\begin{center}
\subfigure[]{
             \includegraphics[scale=0.4]{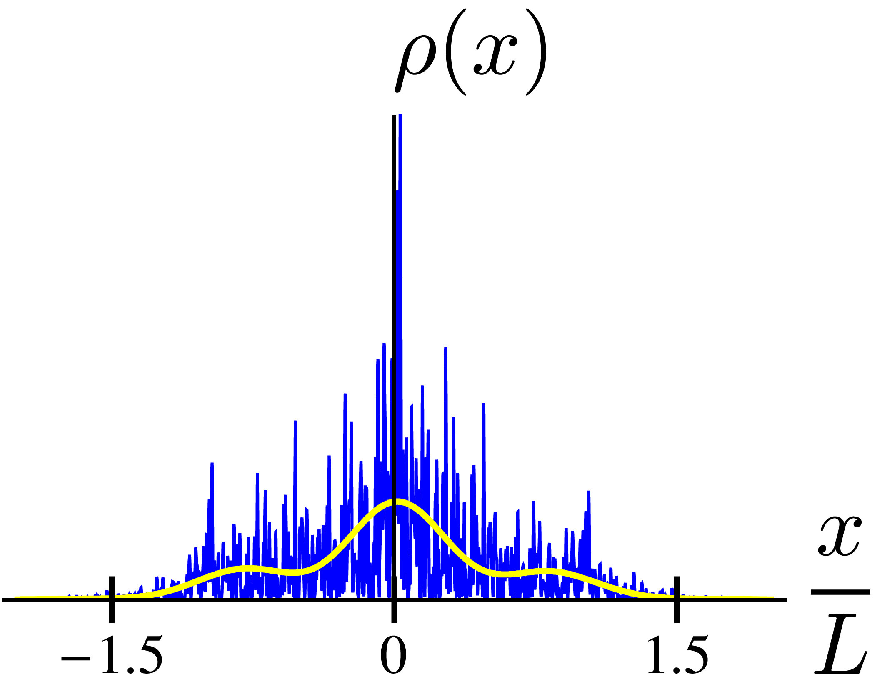}}
\hspace{0.25in}
\subfigure[]{
             \includegraphics[scale=0.4]{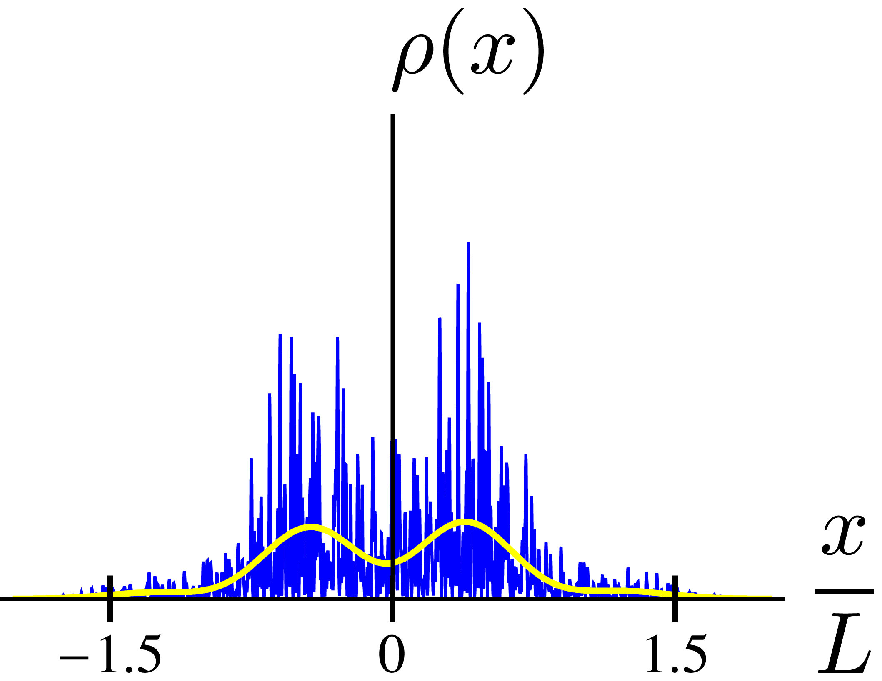}} \\
\hspace{-0.2in}
\subfigure[]{
             \includegraphics[scale=0.38]{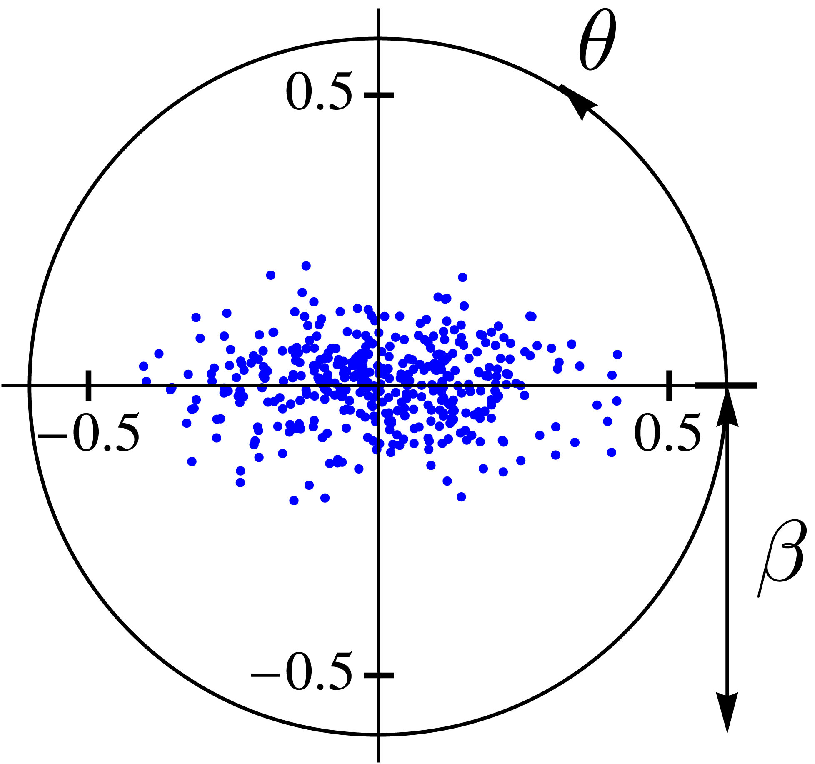}}
\hspace{0.25in}
\subfigure[]{
             \includegraphics[scale=0.43]{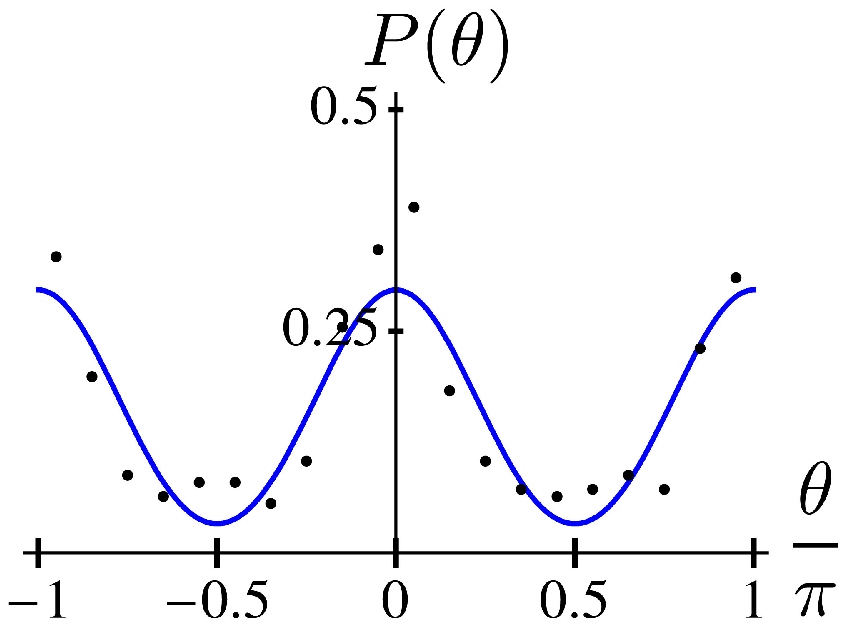}}
\caption{\small{(color online). a),b) Single-shot density profiles (blue solid line) for $\theta=0$ and $\theta=\pi$, respectively. 
The yellow solid line is the fitting function (see text). 
c) polar plot and d) probability phase distribution (dots) after $400$ interference shots.  
The blue solid line in d) is a guide to the eye and the phase values are distributed among 20 bins. 
Here $\nt=1000$ $^{87}$Rb atoms released from a 1D optical lattice, with a time of flight $\tf=20$ ms.
The trapping parameters are $a=5.3$ nm, $d=0.27$ $\mu$m, 
$V_{0}/h=243$ KHz, $\omega_{\perp}=2\pi\times 70$ Hz and $\omega_{x}=2\pi\times 6$ Hz, and
the formation time of the superposition is  $t_{\pi}\sim 60$ ms.} \label{lattice}}
\end{center}
\end{figure}

FIG.~\ref{lattice}(a),(b) show typical single-shot spatial density profiles of 
$\nt=1000$ $^{87}$Rb atoms loaded into $\ns=164$ sites of a 1D optical lattice and squeezing parameter $s=3/2$. 
FIG.~\ref{lattice}(a) corresponds to a fitted value $\theta=0$ while FIG.~\ref{lattice}(b) to $\theta=\pi$. 
In FIG.~\ref{lattice}(c) we plot the polar diagram after $400$ shots. 
Dots are fitting results of repeated simulations: the distance from the 
origin is the amplitude $\beta$, while the polar angle is the phase $\theta$. 
In FIG.~\ref{lattice}(d) we show the phase probability distribution:
the values of $\theta=0$ and $\theta=\pi$ are strongly favored, thus revealing 
the presence of a macroscopic superposition of phase states.
The loss of visibility, compared to the double well case, 
is caused by zero-point energy differences among neighboring sites as will be discussed below. 
In the lattice case faster oscillations with wavelength $L/n$ ( $n$ integer ) can also be fitted out \cite{modulation}. The latter correspond to the combination of relative phases between wells distant $n d$. 
Thus, for instance, phases fitted from two oscillations having distinct wavelengths, can differ by $\pi$ (each fit can provide $0$ or $\pi$ with equal probability).
This would be a clear signature of the creation of the superposition obtained in a single interference experiment. 

{\it Discussion.} The superposition of phase states of a few atoms have probably been 
already created experimentally during the collapse and revival of a matter wave 
field trapped in a three-dimensional optical lattice \cite{Bloch_2002}.
Unfortunately, no signature could have been seen 
in the interference patterns, due 
to column averaging \cite{Bach_2004}.
A one-dimensional configuration is therefore optimal. 
Moreover, in order to avoid dephasing and drifting, the trapping configuration must be such that $\U_{j}$ and $\ej$, respectively, take site independent values.  
In typical experiments the one-dimensional lattice is superimposed to a confining harmonic potential 
with frequencies $\omega_{x}$ along the lattice direction and $\omega_{\perp}$ in 
the transverse directions, which gives $\ej=\Omega (j-N_S/2)^{2}$, with $\Omega=m\omega_{x}^{2}d^{2}/2$. 
Therefore, we require the maximum drift rate between neighboring sites to be much smaller than the creation time of the
phase states superposition, $\Omega N_S/\hbar \ll 1/t_{\pi}$. 
In order to have a site-independent interaction energy, we need the each local chemical potential $\mu_j=N_j\U_j\ll \tilde{\omega_{x}},\omega_{\perp}$, where $\tilde{\omega_{x}}$ is the frequency of each well along the lattice direction.
It is worth to emphasize that these constraints are much relaxed for a double well setup where any asymmetry between the wells would just lead to a global drift of the relative phase distribution, making the creation protocol very robust, while an unbalanced occupation of the wells would only decrease the visibility of the spatial interference fringes.

{\it Decoherence.} Finally, we take into account the possibility for one, two and three-body losses \cite{Sinatra_1998}, which can rapidly destroy the coherent superposition of atomic states.
The condition to make these processes negligible is $\lambda^{[p]}t_{c}<1$ ($p=1,2,3$),
where $t_c$ is the formation time of the macroscopic superposition of phase states,
$\lambda^{[p]}=K^{[p]}\nj^{p}\int d\boldsymbol{r}\pj^{2p}(\boldsymbol{r})$ 
is the loss rate relative to the $p$-body process
and $K^{[p]}$ is an experimentally 
measured rate constant
\cite{nota08}.
The optimal configuration is obtained for a particle density low enough to have only 
a few losses events (usually the three-body losses dominate since they scale as density to the third power), 
but high enough to allow a clear extraction of the phase from the interference 
pattern.
For the lattice setup used in FIG.~\ref{lattice}, we have $\lambda_{0}^{[1]}t_\pi=0.002$, 
and $\lambda_{0}^{[3]}t_\pi=0.0005$, in the central well, with $t_{\pi}\sim 60\ ms$. 
On the other hand, for the experimental double well trap parameters of 
\cite{Oberthaler_2006}, for instance, 
each well should contain no more than $\nj\sim 400$ particles which would give
a formation time $t_{\pi/2}$ approximately equal to the decoherence time, $\sim 500\ ms$. 

{\it Conclusions.} A macroscopic superposition of phase states can be realized experimentally 
with a condensate trapped in a periodic potential. The macroscopic coherence is robust against
asymmetries and decoherence and can be unambiguous detected from the interference patterns 
of the overlapping condensates.

{\it Acknowledgements.}
Discussions with M. Fattori, M. Inguscio, G. Modugno, J. Esteve and G. Watanabe are acknowledged.


\begin{thebibliography}{99}
\bibitem{nota02} We introduce the operators 
$\hat{J}_{x}=(\hat{c}_1^{\dag}\hat{c}_2^{\phantom{\dag}}+\hat{c}_2^{\dag}\hat{c}_1^{\phantom{\dag}})/2$, 
$\hat{J}_{y}=(\hat{c}_1^{\dag}\hat{c}_2^{\phantom{\dag}}-\hat{c}_2^{\dag}\hat{c}_1^{\phantom{\dag}})/2i$ and $\hat{J}_{z}=(\hat{c}_1^{\dag}\hat{c}_1^{\phantom{\dag}}-\hat{c}_2^{\dag}\hat{c}_2^{\phantom{\dag}})/2$, 
which satisfy the SU(2) algebra, with Casimir invariant 
$\hat{\boldsymbol{J}}^{2}=\fracp{\nt}{2}(\fracp{\nt}{2}+1)$.
The decoupled symmetric double well Hamiltonian $\hat{H}_2 \propto \hat{c}_1^{\dag}\hat{c}_1^{\dag}\hat{c}_1^{\phantom{\dag}}\hat{c}_1^{\phantom{\dag}}+
\hat{c}_2^{\dag}\hat{c}_2^{\dag}\hat{c}_2^{\phantom{\dag}}\hat{c}_2^{\phantom{\dag}}$ 
is thus proportional to $\hat{J}_{z}^{2}$, dropping terms proportional to $N_T$.
\bibitem{nota04} The interwell barrier must be raised within a timescale smaller than the inverse of the
trapping frequencies of each well. A residual tunnelling would not prevent the formation of macroscopic
superpositions as long as the correspondent tunnelling rate is smaller than $1/t_{\pi/2}$ (or $1/t_{\pi}$ in the lattice case).
\bibitem{nota100}
A different experimental signature can be obtained by increasing the initial number-squeezing 
so that the tails of the phase distribution of the evolved state can overlap and interfere
(in the case of a mixture the interference is absent). 
Further signatures are provided by the presence of more than two peaks at
intermediate times \cite{nota03}, 
or by the revival of the initial state at $t_{2\pi}$ ($t_{\pi}$) for $N_T$ even (odd).
\bibitem{notamodi}  
Eq.(\ref{superp}) can be rewritten as 
$(e^{-i\fracp{\pi}{4}}(\hat{c}_{x_+}^{\dag})^{\nt}+e^{i\fracp{\pi}{4}}(\hat{c}_{x_-}^{\dag})^{\nt})|0\rangle$ for even and 
$(e^{-i\fracp{\pi}{4}}(\hat{c}_{y_-}^{\dag})^{\nt}+e^{i\fracp{\pi}{4}}(\hat{c}_{y_+}^{\dag})^{\nt})|0\rangle$) odd $N_T$, with
$\hat{c}_{x_\pm}^{\dag}=(\hat{c}_{1}^{\dag}\pm \hat{c}_{2}^{\dag})/\sqrt{2}$, 
$\hat{c}_{y_\pm}^{\dag}=(\hat{c}_{1}^{\dag}\pm i\hat{c}_{2}^{\dag})/\sqrt{2}$
and $|0\rangle$ is the vacuum.
\end{thebibliography}
\end{document}